\documentclass[pdflatex,sn-mathphys-num]{sn-jnl}% Math and Physical Sciences Numbered Reference Style
\usepackage{graphicx}%
\usepackage{multirow}%
\usepackage{amsmath,amssymb,amsfonts}%
\usepackage{amsthm}%
\usepackage{mathrsfs}%
\usepackage[title]{appendix}%
\usepackage{xcolor}%
\usepackage{textcomp}%
\usepackage{manyfoot}%
\usepackage{booktabs}%
\usepackage{algorithm}%
\usepackage{algorithmicx}%
\usepackage{algpseudocode}%
\usepackage{listings}%
\usepackage{hyperref}

\theoremstyle{thmstyleone}%
\newtheorem{theorem}{Theorem}%  meant for continuous numbers
\newtheorem{proposition}[theorem]{Proposition}% 

\theoremstyle{thmstyletwo}%
\newtheorem{example}{Example}%
\newtheorem{remark}{Remark}%

\theoremstyle{thmstylethree}%
\newtheorem{definition}{Definition}%

\begin{document}

\title[Computation of Optimal Type-II Progressing Censoring Scheme Using Genetic Algorithm Approach]{Computation of Optimal Type-II Progressing Censoring Scheme Using Genetic Algorithm Approach}


\author*[1]{\fnm{Ujjwal} \sur{Roy}}\email{ujjwalmath@gmail.com}

\author[2]{\fnm{Ritwik} \sur{Bhattacharya}}%\email{ritwik.bhatta@gmail.com}


\affil*[1]{\orgdiv{Department of Statistics}, \orgname{Central University of Odisha}, \orgaddress{\city{Koraput}, \postcode{763004}, \state{Odisha}, \country{India}}}

\affil[2]{\orgdiv{Department of Mathematical Sciences}, \orgname{University of Texas at El Paso}, \orgaddress{\street{El Paso}, \city{TX}, \country{USA}}}

%\affil[3]{\orgdiv{Department}, \orgname{Organization}, \orgaddress{\street{Street}, \city{City}, \postcode{610101}, \state{State}, \country{Country}}}


\abstract{The experimenter must perform a legitimate search in the entire set of feasible censoring schemes to identify the optimal type II progressive censoring scheme, when applied to a life-testing experiment. Current recommendations are limited to small sample sizes. Exhaustive search strategies are not practically feasible for large sample sizes. This paper proposes a meta-heuristic algorithm based on the genetic algorithm for large sample sizes. The algorithm is found to provide optimal or near-optimal solutions for small sample sizes and large sample sizes. Our suggested optimal criterion is based on the cost function and is scale-invariant for both location-scale and log-location-scale distribution families. To investigate how inaccurate parameter values or cost coefficients may affect the optimal solution, a sensitivity analysis is also taken into account.}




\maketitle

\section{Introduction}\label{sec1}

One of the major disadvantages of the widely used censoring schemes, such as Type-I, Type-II, and hybrid censoring, Balakrishnan and Cramer(2014)\cite{balakrishnan2014art}, Balakrishnan and Aggarwala(2000) \cite{balakrishnan2000progressive}, Bhattacharya et al.(2016) \cite{bhattacharya2016optimum}, do not permit units to be lost or removed from the test at any point other than the experiment's final termination time point. Cohen(1963,1966) \cite{cohen1963progressively}, \cite{cohen1966life} extensively highlighted the importance of removing test units throughout the experiment. Sometimes when removed items can be used for some other tests, intermediate removal may be desirable, or to save time and cost,it might be essential, for instance, to leave the testing facilities for other experiments. Cohen(1963,1966) \cite{cohen1963progressively}, \cite{cohen1966life} introduced a new censoring scheme called the progressive censoring scheme in response to the rigidity of the removal of test units of traditional censoring schemes. Here is how the plan is explained. With $n$ test units, the experiment is started. The number of failures is indicated by an integer $m(\leq n)$ that is fixed beforehand. $R_1$ of the remaining $n-1$ test units are taken out from the experiment at the time of the first failure. In a similar manner, $R_2$ of the remaining $n-R_1-2$ test units are eliminated from the experiment at the time of second failure, and so forth. Finally, all remaining $n-R_1-R_2-\cdots-R_{m-1}-m(=R_m)$ test units are eliminated from the experiment at the time of the $m$th failure. Before an experiment is conducted, the values of $R_i$s are fixed. This is known as the Type-II progressive censoring scheme. Note that when we set $R_1=0,R_2 =0,\cdots,R_{m-1} =0$ and $R_m =n-m$ we get the Type-II censoring scheme. Balakrishnan and Aggarwala(2000) \cite{balakrishnan2000progressive} give more details on progressive censoring.\\

\indent In order to carry out a life-testing experiment under Type-II progressive censoring, the experimenters had to prespecify the values of $n, m$ and $R_i$s. The $R_i$s can be freely chosen for given $n$ and $m$, with a total of ${n-1 \choose m-1}$ possible combinations. Let us denote the set of admissible censoring schemes by
{\footnotesize$$\mbox{CS}(n, m)=\left\{\mathcal{R}=(R_1, R_2,\cdots,R_m)\in \mathcal{N}^m~ |~ \sum_{i=1}^m R_i=n-m,~ \mathcal{N}= \{0, 1,\cdots,n-m\} \right\}.$$}
    
\noindent For moderate $n$ and $m$, the cardinality of the set $\mbox{CS}(n, m)$ is quite large. For example, $|\mbox{CS}(65, 15)|=47855699958816$, where $|\textit{C}|$ is the cardinality of the set $\textit{C}$. It is important to determine the optimal censoring scheme among all those available schemes. A suitable utility function choice, $\xi:\mbox{CS}(n,m)\rightarrow [0, \infty)$, can be maximized or minimized to achieve this. The literature has given a great deal of attention to the selection of $\xi$. For a single parameter case of normal and extreme value distribution, Balakrishnan and Aggarwala (2000) \cite{balakrishnan2000progressive} defined $\xi$ as the variance of the best linear unbiased estimator (BLUE) of the model parameter and illustrated by minimizing it only for small $n$ and $m$ up to $n=50$ and $m= 3$ due to the significant computational complexity. In order to find the optimal solution for generalized Pareto distribution,Burkschat et al. (2006,2007)  \cite{Burkschat2006optimal},\cite{ burkschat2007optimality} employed the same utility function. Also, Burkschat (2008) \cite{Burkschat2008optimality} explored for further choices of $\xi$ as expected test duration, total time on test and variance of the test time. One step progressive censoring, in which censoring occurs only at one failure time i.e. precisely one $R_i$ is positive, has been found to be the optimal censoring scheme in all these works. Let us denote this one-step progressive censoring as
$$\mbox{OSC}(j)=(0,\cdots, 0,\underbrace{n-m}_{\tiny{j\mbox{th position}}}, 0,\cdots, 0), $$ that is, the censoring is carried out at $j$ the failure time of life testing. Note that $\mbox{OSC}(m)$ is the usual Type-II censoring scheme. Balakrishnan et al. (2008) \cite{balakrishnan2008fisher} and Ng et al. (2004) \cite{ng2004optimal} outlined three options for $\xi$ based on the asymptotic variance-covariance matrix and Fisher information, and they produced the optimal censoring schemes for the Weibull distribution. In order to determine the best schemes for the Birnbaum-Saunders and generalized exponential distributions, Pradhan and Kundu (2009,2013) \cite{pradhan2009progressively},\cite{ pradhan2013inference} presented a criterion based on the asymptotic variance of the estimated $p$th quantile of the underlying lifetime distribution. It should be noted that the aforementioned works have examined the process of determining the optimal censoring schemes for small values of $n$ and $m$. Recently, Bhattacharya et al. (2014,2016) \cite{bhattacharya2014optimum},\cite{bhattacharya2016optimum} proposed an optimal criterion based on the cost function and, using the variable neighborhood search (VNS) algorithm, obtained the optimal scheme for upto $n=45$ and $m=15$. Balakrishnan and Cramer(2014) \cite{balakrishnan2014art} provide a brief overview of the literature on the topic of optimal progressive censoring schemes.\\

\indent In this work, we propose a meta-heuristic approach based on the genetic algorithm (GA) to obtain the optimal censoring scheme for moderately large values of $n$ and $m$. We also propose a new choice of $\xi$ based on the total cost associated with a life-testing experiment that is used by Bhattacharya et al. (2016) \cite{bhattacharya2016optimum}. To the best of our knowledge, no effective algorithm exist to obtain the optimal censoring scheme for large values of $n$ and $m$. The suggested algorithm is incredibly easy to construct and use. It is observed that the proposed algorithm GA, gives better result as compared to existing algorithm (i.e. VNS), for small $n$ and $m$. Moreover, the proposed algorithm provides the optimal or near-optimal solution for moderately large values of $n$ and $m$ within a reasonable computational burden, which may be preferred to the solution obtained through VNS.\\

\indent The rest of the paper is organized as follows. In Section \ref{sec2}, we outline the progressively censored data and associated findings that are needed to compute the optimal schemes. In Section \ref{sec3}, we present an optimality criterion based on the cost function. The foundations of GA and the development of the proposed algorithm are discussed in Section \ref{sec4}. Section \ref{sec5} provides the optimal schemes obtained using the cost minimization-based optimal criterion used in Section \ref{sec3}. Finally, some concluding remarks are made in Section \ref{sec6}.

\section{Progressively censored data}\label{sec2}
	\paragraph{}
	Let $Y_i;1\leq i \leq n$ be the i.i.d. lifetimes of $n$ units on a life-testing experiment with a common distribution function $F(y; \phi)$ and a density function $f(y; \phi)$, where $\phi\in\Phi$ is an open subset of $\mathbb{R}^q$. The number of failures $m$ is fixed before the experiment starts and a censoring scheme $R_i;1\leq i \leq m$, satisfying equation $\sum_{i=1}^m R_i=n-m$ with pre-specified $R_i\geq0$. We denote the corresponding failure times as $Y_{1:m:n}\leq Y_{2:m:n}\leq \cdots\leq Y_{m:m:n}$ and is different from the usual $i$th order statistic from a sample of size $n$, $Y_{i:n}$. $Y_{i:m:n}$ is known as progressively Type-II censored ordered statistics (see Balakrishnan and Aggarwala 2000) \cite{balakrishnan2000progressive}. On the basis of the progressively censored data $Y_{i:m:n}= y_{i:m:n} (1\leq i\leq m)$, the likelihood function can be written as 
	\begin{equation}\label{likelihoodfunc}
		L(\phi) = \kappa^{'}\prod_{i=1}^{m}f(y_{i:m:n}; \phi)\{1-F(y_{i:m:n}; \phi)\}^{R_i},
	\end{equation}
	where $\kappa^{'}$ is normalizing constant, $\kappa^{'}=n(n-R_1-1)(n-R_1-R_2-2)\cdots(n-R_1-R_2-\cdots-R_{m-1}-m)$ and $y_{i:m:n}$ is the observed value of $Y_{i:m:n}$. The corresponding Fisher information matrix for the vector parameter $\phi$, 
    is given by \begin{equation} \label{FI1}
		\mathcal{I}(\phi)= -E\left[\frac{\partial^2}{\partial\phi\partial\phi^{T}}\ln L(\phi)\right].
	\end{equation}

    \noindent
	A direct computation formula for Fisher information matrix given by Dahmen et al. (2012) \cite{dahmen2012and} as follows
	\begin{equation}\label{FI_Pro}
		\mathcal{I}(\phi)= \int_{-\infty}^{\infty} \Big \langle \frac{\partial}{\partial \boldsymbol \phi}\ln h(y; \phi)   \Big \rangle \sum_{i=1}^m f_{Y_{i:m:n}}(y; \phi) \mbox{d}y,
	\end{equation}
	where $f_{Y_{i:m:n}}(y; \phi)$ is the density function of $Y_{i:m:n}$, $h(y; \phi)$ is the hazard rate of $Y$ and $(\frac{\partial}{\partial \boldsymbol \phi})\ln h(y, \phi)=(\frac{\partial}{\partial \phi_1}\ln h(y, \phi),\frac{\partial}{\partial \phi_2}\ln h(y, \phi), \cdots,\frac{\partial}{\partial \phi_q}\ln h(y, \phi))$, $ \boldsymbol \phi=(\phi_1, \phi_2,$ $\cdots,\phi_q)$. Here, $\langle \cdot \rangle$ is defined as the matrix $\langle a \rangle=a.a^{T}$, $a\in \mathbb{R}^q$. There are different representations for the density function of $Y_{i:m:n}$. In our study, we use the Camp-Cramer representation (see Balakrishnan 2007) \cite{balakrishnan2007progressive} and it is given by
	\begin{equation}\label{pdfTimn}
		f_{Y_{i:m:n}}(y; \phi)= \sigma_{i-1}\sum_{k=1}^i a_{k,i}\{1-F(y; \phi)\}^{\gamma_k-1}f(y; \phi), \mbox{~for $i=1,2,...,m$,}
	\end{equation}
	where $\gamma_r = m-r+1+\sum_{i=r}^m R_i,$ for $r=1,2,...,m$; $\sigma_{r-1}= \prod_{i=1}^r \gamma_i,$ $a_{i,r}= \prod_{\substack{j=1\\ j\neq i}}^r \frac{1}{\gamma_j-\gamma_i},$ for $1\leq i \leq r \leq m $ with $a_{1, 1}=1$.

\section{Optimal criteria based on cost               minimization}\label{sec3}
	\paragraph{}
	The two most often used optimality criteria in statistical design problems are the A- and D-optimality. Under the A-optimality criterion, the utility function $\xi$ is the sum of variances of the estimated model parameters, while the D-optimality criterion uses an overall major of variability given by the determinant of the co-variance matrices of the estimated model parameters (Liski et al. 2002) \cite{liski2002topics}. In the context of Type- II progressive censoring scheme, Ng et al. (2004), Balakrishnan et al. (2008) and Dahmen et al. (2012) \cite{ng2004optimal}, \cite{balakrishnan2008fisher}, \cite{dahmen2012and} consider A- and D- optimal criteria for Weibull, Lomax, normal, and extreme value distributions. Pradhan and Kundu (2009,2013) \cite{pradhan2009progressively},\cite{pradhan2013inference} considered an optimality criterion based on estimated pth quantile of the distribution of lifetime $Y$, instead of using conventional A- and D- optimal criteria. Specifically, by integrating over $p$, they considered  $\xi(\mathcal{R})= \int_{0}^{1}\mbox{Var}[\ln\hat{Y}_p]\mbox{d}p, \mathcal{R}\in \mbox{CS}(n, m)$, where $\hat{Y}_p$ is the maximum likelihood estimate of $p$th quantile of lifetime distribution of $Y$. It should be noted that the integral over $[0,1]$ represents the aggregate variance of quantile estimates over all quantile points, so that $\xi$ is independent of the choice of $p$.

    \noindent For the purpose of identifying the optimal censoring schemes, the majority of the works used maximum information as their optimality criterion. Few studies, nevertheless, have looked at minimizing the experiment's cost as the optimality criterion. The total cost of a life-testing experiment is a crucial consideration in an industrial setting. The cost function was taken into account under a Type-II censoring scheme by Epstein (1960) and Blight (1972) \cite{epstein1960sampling},\cite{blight1972most}, while Ebrahimi (1988) and Bhattacharya et al. (2014) \cite{ebrahimi1988determining},\cite{bhattacharya2014optimum} considered a similar cost function under a hybrid censoring scheme. Recently, Bhattacharya et al. (2016) \cite{bhattacharya2016optimum} also considered similar cost functions under a Type-II progressing censoring scheme. They have shown that the optimal solution obtained by minimizing their cost function is scale invariant in the sense that the optimal scheme remains unaltered if lifetime is multiplied by some constant that is, lifetime is measured in some other unit. They have also shown that the A- and D- optimal schemes are not scale invariant. In this work, we proposed an optimality criterion with an experiment. Following the idea of Bhattacharya et al. (2014) \cite{bhattacharya2014optimum}, we introduce a cost function which is total cost associated with the life testing experiment under progressive censoring as follows
	\begin{equation}\label{costpsi}
		\xi(\mathcal{R})=\kappa_1m+\kappa_2E[Y_{m:m:n}]+\kappa_3\int_0^1\mbox{ Var}[\ln \hat {Y}_p]\mbox{d}p, \mathcal{R}\in \mbox{CS}(n, m),
	\end{equation}
	where $\kappa_1$, $\kappa_2$ and $\kappa_3$ are cost per unit of failures, cost per unit of time that the experiment is being conducted and cost of imprecision (variance) of the estimates of the unknown parameters of the lifetime distribution under consideration, respectively. Note that $Y_{m:m:n}$ is the random variable representing the duration of the life testing experiment.\newline

    \noindent \textbf{Result:} The cost function defined in Eq. (\ref{costpsi}) is scale invariant for location-scale and log-location-scale family of distributions.\newline
 
	\noindent \textbf{Proof:} To prove the stated result, let us consider the transformation $Y^*=wY$ of the lifetime random variable. Let us define $Y^*_{m:m:n}$, $\hat {Y}^*_p$, $\kappa^*_1$, $\kappa^*_2$ and $\kappa^*_3$ as earlier with self explanatory notation in transformed lifetime $Y^*$. Now it is easy to verify that $E[Y^*_{m:m:n}]=wE[Y_{m:m:n}]$. Also $\kappa^*_1=\kappa_1$, $\kappa^*_2=\kappa_2/w$ and $\kappa^*_3=\kappa_3$. Therefore, we only need to show the invariance property of the variance measure $\int_0^1\mbox{ Var}[\ln \hat {Y}_p]\mbox{d}p$ which can be proved if the linear relationship $\hat {Y}^*_p=w\hat {Y}_p$ holds. This relationship holds good for location-scale and log-location-scale family of distributions(see Bhattacharya et al. 2014) \cite{bhattacharya2014optimum}. Hence, the  proof.\newline

\section{Genetic Algorithm}\label{sec4}
	\paragraph{}
	To develop optimal censoring schemes, one needs to choose a utility function $\xi$ as discussed in the preceding section, and therefore the optimization problem can be formulated as follows
	\begin{eqnarray}\label{utility}
		&&\underset{\mathcal{R}\in \tiny{\mbox{CS}}(n, m)}{\text{minimize or maximize}}~ \xi(\mathcal{R})
	\end{eqnarray}
	subject to constraints, if any. This is simply a discrete optimization problem in which the solution space consists of a finite number of $m$-tuples (of $R_i$'s) with integer elements, and our job is to find that set of $R_i$'s which minimize/maximize $\xi$. For small $n$ and $m$, the optimal solution can be obtained by exhaustive search (see Ng et al. 2004; Balakrishnan et al. 2008; Pradhan and Kundu 2009,2013) \cite{ng2004optimal},\cite{balakrishnan2008fisher},\cite{pradhan2009progressively},\cite{pradhan2013inference}. For large values of $n$ and $m$, the computation of an optimal solution by exhaustive search is really challenging with respect to computational resources, since the cardinality of the set $\mbox{CS}(n, m)$ is too large. The biggest disadvantage of the VNS is that (i) the efficacy is contingent upon the selection of neighborhood structures and the sequence of their application, (ii) it may necessitate problem-specific expertise to delineate successful neighborhood arrangements, and (iii) the search technique may incur significant computational costs, particularly for extensive issue cases or intricate neighborhood configurations. To the best of our knowledge, there is no efficient algorithm in the literature to address this issue. In this work, we propose a meta-heuristic approach to obtain the optimal solution using the Genetic Algorithm (GA). GA is a highly regarded learning algorithm that does not rely on derivatives and does not depend on the initial solutions.\newline

    GA is an evolutionary stochastic optimization technique. GA have extensive applications across various fields, introduced by Holland in the 1960s and subsequently advanced by Goldberg. GAs are stochastic search methods that differ from traditional optimization techniques in several characteristics. The following is a summary (see Goldberg and Holland 1988) \cite{Goldberg1988GeneticAA}: GA employs the coding of solutions rather than the values of the solutions. GA seek a population of points rather than individual points at each iteration to attain the optimal solution. GA evaluates only the objective function to achieve the optimal solution, without requiring other information such as the derivatives of the objective function (see Miguel and Sahinidis 2013) \cite{rios2013derivative}. This feature enables the identification of the optimal solution for non-derivative objective functions. GA employ probabilistic transition rules rather than deterministic ones. In each iteration, every candidate solution in the population is represented by a chromosome. GA seeks to achieve the optimal solution by employing genetic operators on chromosomes and assessing them through the fitness (objective) function. In this work, we consider real-coded genetic algorithms (GAs) that do not use any coding of the problem variables, instead they work directly with the variables. Due to its foundation in natural selection and genetic principles, GA employs genetic operators known as selection, crossover, and mutation. The Tournament Selection Process as the Selection Operator, Blend Crossover as the Crossover Operator, Uniform Random Mutation as Mutation Operator, and this can be implemented in a genetic algorithm as follows.\newline

    \textbf{Tournament Selection Process}:
    In this process, we randomly select a subset of individuals from the population. The size of this subset is called the tournament size($k$). Evaluate the fitness of each individual in the subset. Choose the individual with the highest fitness from the subset as a parent. Repeat the process until the desired number of parents is selected. For example: Consider a population of individuals with their fitness values:
    \begin{itemize}
        \item 	Individual A: Fitness = 2.3
        \item 	Individual B: Fitness = 1.9
        \item 	Individual C: Fitness = 3.2
        \item 	Individual D: Fitness = 5.8
        \item 	Individual E: Fitness = 7.4
    \end{itemize}
    Assume that we are using a tournament size of 4 (k=4) and we need to select 2 parents. We randomly select 4 individuals: A, B, D, E for the first tournament and select the individual E with the highest fitness=7.4. For the second tournament, we randomly select 4 individuals: A, B, C, D and select individual D with highest fitness=5.8. So, the selected parents for reproduction are Individual D and Individual E.\newline

    \textbf{Blend Crossover}:
    The Crossover operation will be executed on a chromosome as determined by the specified crossover rate (Cr). BLX-$\alpha$(Blend Crossover) was proposed by Eshelman and Schaffer in 1993 \cite{eshelman1993real}. Suppose that we have two parents ($P_1$) and ($P_2$) who participate in the crossover, then the children solutions are as follows:\newline
    $C_1=(1-\gamma)*P_1+\gamma*P_2$ \newline
    $C_2=(1-\gamma)*P_2+\gamma*P_1$ \newline
    where $\gamma=(1+2\alpha)*$r$-\alpha$ and $\alpha$ is a fixed value and $r$ is a random number between $(0.0,1.0)$. For example: If $P_1$ =10.53 and $P_2$ =15.39, $r$=0.4, $\alpha$=0.5 and $\gamma$=0.3.\newline
    Then $C_1$=11.988 and $C_2$=13.932.\newline

    \textbf{Uniform random mutation}:
    Introduces diversity into the population by randomly altering the value of a gene within its predefined range. We randomly choose a gene ($a_i$) within the range $([x_i,y_i])$ of the chromosome with a predefined mutation rate (Mr). Assign a new value to ($a_i$) by sampling uniformly from the range $([x_i,y_i]):[a_i=U(x_i,y_i)]$ where$(U(x_i,y_i))$denotes a uniform random number within the range$([x_i,y_i])$. For example: Let us consider an example with a chromosome represented by three genes ([3.4,2.6,4.5]) and ranges for each gene ([2.0,4.0]),([1.0,5.0]),([2.0,6.0]) respectively. Suppose we randomly select the third gene $(a_3=4.5)$. The range for $(a_3)$ is ([2.0,6.0]). Generate a new value for $(a_3)$ uniformly within ([2.0,6.0]). Let us say the random value generated is 4.1. Then,the mutated chromosome would be:
    \begin{itemize}
        \item Original Chromosome: ([3.4,2.6,4.5])
        \item Mutated Chromosome: ([3.4,2.6,4.1])
    \end{itemize}
    This mutation introduces variability into the population, helping the algorithm explore the search space more effectively.\newline

    \subsection{Algorithm to obtain optimal schemes}\label{sec4.1}
    \paragraph{}
    GA Steps for parameter estimation of Type-II progressive censoring from the Weibulll distribution are as follows:\newline

    Let us generate an initial population of (N) individuals, each represented by a vector of real numbers. Then evaluate the fitness of each individual in the population. Select parents from the population according to their fitness using tournament selection. Apply the BLX-$\alpha$ crossover operator with crossover probability 0.8 to pairs of parents to generate offspring. Apply the uniform random mutation with mutation probability 0.1 to the offspring to introduce variability. Evaluate the fitness of the offspring. Replace the current population with the new offspring (or use elitism to retain the best individuals). Check if the termination condition is met (for example, a maximum number of generations or a satisfactory fitness level). If not, then repeat selection, crossover, and mutation. Return the best individual found as the solution. The pseudo code of the algorithm is described as follows:\newline

    \vspace{5pt}
    Initialization. Choose an initial solution $\mathcal{R}_{0}\in CS(n,m)$ and the stopping condition if no further improved solution is found within its maximum possible iteration and take it as the optimal solution $\mathcal{R}_{opt}$, where $\mathcal{R}_{0}= (R_{01},R_{02},...,R_{0m})$. \newline
    Repeat. Until the stopping condition is satisfied, repeat the following steps: 
    \begin{enumerate}
        \item Select parents for the next generation using the tournament selection procedure from the population.
        \item Apply BLX-$\alpha$ Crossover operator for diversity in population.
        \item Use uniform random mutation to avoid the local optimal problem.
        \item Calculate fitness for the new population.
    \end{enumerate}

\section{Optimal scheme based on cost minimization}\label{sec5}
	\paragraph{}
	In this section, we consider different choices of $\xi(\mathcal{R})$ as utility function given by Eq.(\ref{costpsi}) and obtain optimal schemes using the Genetic Algorithm. 
	\subsection{Cost minimization based optimal designs for known $n$ and $m$}
	\paragraph{}
	We consider that the lifetime $Y$ follows the Weibull distribution with distribution function given by 
	\begin{equation}\label{Wcdf}
		F(y; \zeta, \rho)=1-e^{-(\rho y)^{\zeta}},~~ \rho, \zeta, y >0,
	\end{equation}
	where $\zeta$ and $\rho$ are shape and scale parameters, respectively. To compute $\int_0^1 Var[\ln \hat {Y}_p]dp$, we need to calculate Fisher information matrix $\mathcal{I}(\phi)$ for vector parameter $\phi=(\zeta, \rho)$  which directly follows from Eq.(\ref{FI_Pro}) as 
	$$\mathcal{I}(\phi)=\begin{bmatrix}\mathcal{I}_{11}(\phi) && \mathcal{I}_{12}(\phi) \\ \mathcal{I}_{21}(\phi) && \mathcal{I}_{22}(\phi) \end{bmatrix},$$ where
	\begin{eqnarray}\nonumber
		\mathcal{I}_{11}(\phi) &=& \frac{1}{\zeta^2} \sum_{i=1}^m  \sum_{k=1}^i \sigma_{i-1} \frac{a_{k, i}}{\gamma_k} \int_0^{\infty} \left(  1+ \ln\left(\frac{z}{\gamma_k}\right)  \right)^2 e^{-z}\mbox{d}z, \\\nonumber
		\mathcal{I}_{22}(\phi) &=& \left( \frac{\zeta}{\rho}  \right)^2 \sum_{i=1}^m  \sum_{k=1}^i \sigma_{i-1} \frac{a_{k, i}}{\gamma_k}, \\\nonumber
		\mathcal{I}_{12}(\phi)~ = ~\mathcal{I}_{21}(\phi) &=& \frac{1}{\rho} \sum_{i=1}^m  \sum_{k=1}^i \sigma_{i-1} \frac{a_{k, i}}{\gamma_k} \int_0^{\infty} \left(  1+ \ln\left(\frac{z}{\gamma_k}\right)  \right)e^{-z} \mbox{d}z.
	\end{eqnarray}
	By delta method, the asymptotic expression of $\mbox{Var}[\ln \hat {Y}_p]$ can be seen to be as $\mbox{Var}[\ln \hat {Y}_p]=(\mathcal{I}^{11}(\phi)/\zeta^4) (g(p))^2+(2\mathcal{I}^{12}(\phi)/\zeta^2\rho )g(p)+(\mathcal{I}^{22}(\phi)/\rho^2)$, where $g(p)=\ln\{ -\ln(1-p) \}$ and, $\mathcal{I}^{11}(\phi)$, $\mathcal{I}^{12}(\phi)$ and $\mathcal{I}^{22}(\phi)$ are the elements of inverse of Fisher information matrix given by $$\mathcal{I}^{-1}(\phi)=\begin{bmatrix}\mathcal{I}^{11}(\phi) && \mathcal{I}^{12}(\phi) \\ \mathcal{I}^{21}(\phi) && \mathcal{I}^{22}(\phi) \end{bmatrix}.$$  Therefore,
	\begin{equation}\nonumber
		\int_0^1 \mbox{Var}[\ln \hat {Y}_p]\mbox{d}p=\frac{\mathcal{I}^{11}(\phi)}{\zeta^4}\int_0^1 (g(p))^2 \mbox{d}p+\frac{2\mathcal{I}^{12}(\phi)}{\zeta^2\rho}\int_0^1 g(p)\mbox{d}p+\frac{\mathcal{I}^{22}(\phi)}{\rho^2}.
	\end{equation}
	\noindent Using the distribution of $Y_{i:m:n}$, $E[Y_{m:m:n}]$ can be computed as
	\begin{equation}\nonumber
		E[Y_{m:m:n}] = \frac{1}{\rho} \Gamma\left(1+\frac{1}{\zeta}\right) \sigma_{m-1}\sum_{k=1}^m \frac{a_{k,m}}{\gamma^{1+\frac{1}{\zeta}}_k} .
	\end{equation}
	Thus, the expressions of the components of $\xi(\mathcal{R})$ in Eq. (\ref{costpsi}) are obtained. To illustrate numerically, we employ the real coded GA of Section \ref{sec4.1} to determine the optimal scheme for various values of n and m. We used various parameter values of the Weibull distribution, specifically, $(\zeta, \rho)=(2, 1), (1, 1)\mbox{~and~}(0.5, 1)$. Table \ref{table:1} presents a comparison between the GA-derived solution and the solution obtained using the VNS algorithm or an exhaustive search in $\mbox{CS}(n, m)$ for small values of $n$ and $m$. The optimal progressive censoring scheme is represented by $\mathcal{R}^*$, with the associated optimal cost given as $\xi(\mathcal{R}^*)$ for both techniques. In the table, $a^*b$ signifies that $a$ is repeated consecutively $b$ times; for instance, $(0^*3, 5)$ denotes $(0, 0, 0, 5)$. The GA produces the relatively better optimal solution as that derived from VNS algorithm and exhaustive search in $\mbox{CS}(n, m)$. For $n=15,m=5$, the parameter values for the two rows in Table \ref{table:1} pertain to the examination of invariance under scale transformation, where the first row denotes the original lifetime $Y$ and the second row signifies the lifetime multiplied by 2. When $Y$ adheres to a Weibull distribution characterised by shape and scale parameters $\zeta$ and $\rho$, respectively, the transformed lifespan $Y^*=wY$ also conforms to a Weibull distribution with the identical shape parameter $\zeta$  and a scale parameter $\rho/w$. The cost coefficients $\kappa_1$ and $\kappa_3$ stay unchanged, but $\kappa_2$ is modified to $\kappa_2/w$ under scale translation. The optimal scheme $\mathcal{R}^*$ and the optimal cost $\xi(\mathcal{R}^*)$ remain unchanged under scale transformation. All computations are executed using the Intel Core i7-1165G7 CPU operating at 2.80 GHz.\newline

    \indent To compute the optimal schemes in tables \ref{table:1}, \ref{table:2}, \ref{table:3}, \ref{table:4}, we have taken initial solution $\mathcal{R}_0$ using random generation from the multivariate hyper-geometric distribution. Ultimately, we derive several optimal strategies for comparatively large values of n and m. We select n = 50, 55, 60, and 65, together with various values of $m (< n)$, utilizing the cost coefficients $\kappa_1 = 10, \kappa_2 = 50,$ and $\kappa_3 = 250$. The optimal schemes are presented in Table \ref{table:2}. It is important to recognize that the optimal solution is not OSC(j) type, particularly for all values of n and m. The optimal cost increases with m for a constant n, as expected. Interestingly, given a fixed m, the optimal cost decreases as n increases. This may be attributed to the fact that $E[Tm:m:n]$ decreases with increasing n, for a constant m (see Theorem 4.1 in Burkschat 2008) \cite{Burkschat2008optimality}, and due to more available information, the expense of imprecision, the third element in Eq. (\ref{costpsi}), also decreases with n. For fixed n and m, the optimal cost of a constant $\rho$ is observed to increase as $\zeta$ decreases, likely due to the increased average failure time with a reduction in $\zeta$. \newline

    \indent The optimal censoring scheme is contingent upon the selection of model parameters of the underlying distribution and the associated cost coefficients. Consequently, in practice, the experimenter must predefine these parameter values and cost factors to achieve the optimal design. Therefore, a sensitivity analysis is necessary to examine the impact of misspecification of the parameter value or the cost coefficient on the optimal solution. Initially, we examine the sensitivity analysis concerning the parameters of the underlying distribution for a fixed set of cost coefficients. We define the relative efficiency of a set of parameter values $\phi$ compared to the set of true parameter values $\phi_0$ based on the cost function as
    \begin{eqnarray*}
	\mbox{RE}_1(\phi)=\frac{E_{\phi_0}[\mbox{Cost for optimal solution under } \phi_0]} {E_{\phi_0}[\mbox{Cost for optimal solution under } \phi]}.
    \end{eqnarray*}
    
    The value of $\mbox{RE}_1(\phi)$ ranges from 0 to 1, with values approaching 1 indicating less sensitivity of the optimal solution to the misspecification of model parameters. It is important to note that due to the invariance property of the cost function, the scale parameter $\rho$ may be fixed at 1, allowing a sensitivity analysis of the Weibull distribution with respect to the shape parameter $\zeta$ and the parameter $\phi$, in general. In this case, we examine $\kappa_1 = 10, \kappa_2 = 50, \kappa_3 = 250$, and $(n, m) = (20, 5)$ for demonstration purposes. The results presented in Table \ref{table:3} demonstrate that the optimal scheme exhibits reduced sensitivity to misspecification of the shape parameter of the Weibull distribution.

    \indent Subsequently, we examine the sensitivity analysis concerning the various cost coefficients. As earlier, for fixed value of $\phi$, we define the relative efficiency of a set of cost coefficients $\mathcal{C}$ in comparison to the true cost coefficients $\mathcal{C}_0$ as
    \begin{eqnarray*}
    \mbox{RE}_2(\mathcal{C})=\frac{E_{\mathcal{C}_0}[\mbox{Cost for optimal solution under }  \mathcal{C}_0]}{E_{\mathcal{C}_0}[\mbox{Cost for optimal solution under } \mathcal{C}]}.
    \end{eqnarray*}
     it has been found that the optimal schemes are also less sensitive to the mis-specification of cost components $\kappa_1$, $\kappa_2$ and $\kappa_3$.

     \subsection{Optimal progressive censoring schemes for known $n$ only}
     Sometimes, the experimenter may be aware of the available sample size $n$ for life testing, necessitating a judgment on the selection of $m$ and $R_i$s under progressive censoring. The optimization problem can be articulated to minimize $\xi(\mathcal{R})$ in Eq. (\ref{costpsi}) with respect to $m$ and $\mathcal{R}$.
      For a given value of $n$, the allowed values of $m$ are $1, 2,\cdots, n$. For every combination of $n$ and $m$, the optimal selection of $R_i$'s in $CS(n, m)$ can be derived using GA as described in Section \ref{sec4.1}. Subsequently, from all potential combinations of $n$, $m$, and the related optimal $R_i$'s, we select the combination that yields the lowest cost as the optimal solution. Furthermore, observe that the optimal solution derived from minimizing the aforementioned cost function is scale invariant. To illustrate, we examine the Weibull distributions with parameters $(\zeta,\rho)=(2, 1), (1, 1) \mbox{~and~} (0.5, 1)$, together with the cost coefficients $(\kappa_1, \kappa_2, \kappa_3)=~(10, 50, 250)$. Table \ref{table:5} presents the optimal schemes along with the optimal selection of $m$ (designated as $m^{*}$) for different values of $n$. The scale-invariant characteristic of the optimal scheme is also demonstrated, as noted in Section \ref{sec3}.
      
\section{Concluding remarks}\label{sec6}
    This study presents GA that surpasses the exhaustive search approach and the VNS algorithm by efficiently identifying near-optimal solutions for small values as well as large values of n and m while minimizing computational burden. The proposed algorithm can be extended to the other variants of progressive censoring schemes with suitable adjustments. The present work goes in that direction and will be present in future papers.

\newpage
\begin{table}[h] 
    \centering
    \caption{Comparison of optimal solutions obtained through the proposed GA with optimal solutions obtained through the VNS algorithm in $\mbox{CS}(n,m)$ for small $n$ and $m$ with $(\kappa_1,\kappa_2,\kappa_3)=(10,50,250)$.}
    \label{table:1}
    \vspace{5pt}
    \begin{tabular}{c c c}
    \toprule
    Parameters($\zeta,\rho$)   & Optimal Solution(VNS) & Optimal Solution(GA) \\ 
       &$\mathcal{R}^*,\xi(\mathcal{R}^*)$ \footnotemark{} &$\mathcal{R}^*,\xi(\mathcal{R}^*)$ \\
    \midrule
        $n=15,m=5$ &  & \\
        (2,1) & (0*4,10),110.433          & (3*2,0*2,4),107.0203  \\ 
        (2,0.5) & (0*4,10),110.433        & (3*2,0*2,4),107.0203  \\
        (1,1) & (0,5,0*2,5),183.9206       & (4*2,0*2,2),173.5582  \\
        (1,0.5) & (0,5,0*2,5),183.9206    &  (4*2,0*2,2),173.5582  \\
        (0.5,1) & (0,7,0*2,3),476.177      & (5,4,0*2,1),438.9402  \\
        (0.5,0.5)& (0,7,0*2,3),476.177      & (5,4,0*2,1),438.9402  \\
        $n=20,m=5$ &  & \\
        (2,1) & (4,0*3,11),108.727          & (6*2,0*2,3),106.4338  \\ 
        (1,1) & (10,0*3,5),177.876          & (7,6,0*2,2),171.006    \\
        (0.5,1) & (0,13,0*2,2),455.426          & (7*2,0*2,1),427.8285  \\
        $n=30,m=5$ &  & \\
        (2,1) & (14,0*3,11),106.822          & (17,0*3,8),105.5198 \\ 
        (1,1) & (20,0*3,5),171.218          & (12,11,0*2,2),168.5684  \\
        (0.5,1) & (22,0*3,3),430.484          & (12*2,0*2,1),418.0383  \\
        $n=30,m=25$ &  & \\
        (2,1) & (0*24,5),321.309          & (0*20,1*5),296.9331  \\ 
        (1,1) & (0*24,5),360.787 & (0*6,1,0*3,1,0*5,1,0*4,1,0*2,1),297.5168 \\
        (0.5,1) & (0*24,5),504.732 & (0*7,1,0*3,1,0*3,1,0*5,1,0*2,1),364.9231 \\
        $n=35,m=10$ &  & \\
        (2,1) & (0*9,25),145.087          & (6*3,0*5,2,5),141.2055  \\ 
        (1,1) & (0,10,0*7,15),180.004     & (7,6,7,0*6,5),170.4071  \\
        (0.5,1) & (0,16,0*7,9),325.050    & (7*2,6,0*6,5),299.73  \\
        $n=40,m=10$ &  & \\
        (2,1) & (0*9,30),144.046          & (7*3,0*5,3,6),141.2033 \\ 
        (1,1) & (0,15,0*7,15),178.464     & (8,9*2,0,1,0*4,3),172.825  \\
        (0.5,1) & (0,21,0*7,9),319.549    & (9*3,0*4,1,0,2),300.8225  \\
        $n=45,m=15$ &  & \\
        (2,1) & (1,0*12,1,28),190.913     & (4*5,0*2,1,0*4,1,4,4),185.4312  \\ 
        (1,1) & (1,0*13,29),206.954       & (6,5,6,3,2,1*2,0*3,1,0*3,5),202.4484 \\
        (0.5,1) & (1,0*13,29),305.052     & (6,5,6*2,3,1,0*6,1,0,2),296.3334 \\ 
        \bottomrule
    \end{tabular}
\end{table}

\footnotetext{The data of the second column refer to Bhattacharya et al. (2016) \cite{bhattacharya2016optimum} }
\begin{sidewaystable}[h] 
    \centering
    \caption{Optimal progressive censoring schemes under Weibull distribution with $(\kappa_1,\kappa_2,\kappa_3)=(10,50,250)$.}
    \label{table:2}
    \vspace{5pt}
    \begin{tabular}{c c c c c}
        \toprule
        $n$ & $m$ & Optimal Solution & Optimal Solution & Optimal Solution  \\ 
          & & $\zeta=2,\rho=1$ & $\zeta=1,\rho=1$ & $\zeta=0.5,\rho=1$  \\
          &    & $\mathcal{R}^*,\xi(\mathcal{R}^*)$&$\mathcal{R}^*,\xi(\mathcal{R}^*)$ & $\mathcal{R}^*,\xi(\mathcal{R}^*)$ \\ 
          & &            &               &                 \\
          \midrule
          50&5&(35,0,3,0,7),105.001&(41,0,0,1,3),166.1668& 
            (42,0*3,3),406.7854 \\
            &10&(10*3,1,0,1,0*3,8),140.4117&(12*3,0*6,4),172.2479&(12,13*2,0*6,2),297.0889  \\
            &15&(4,5,1,2*2,1,2,0,1,2*2,0,2,7,4),154.6625&(6,1,8,1*7,0*3,1,12),205.7382&
            (14,5,1,2,1*2,2,1,0,1*3,0,1,4),304.6833 \\
         
          55&5&(40,0*3,10),103.8156&(45,0*3,5),163.2054& 
            (47,0*3,3),401.9119  \\
            &10&(13*3,0*6,6),128.3827&(14,13,12,1,0*5,5),166.0358& 
            (14*2,15,0*6,2),293.8523 \\
            &15&(5,4,5,4,5,2,1*5,0,1,4,5),185.1868&(4,9,5,6,2*4,1*3,0*2,1,4),196.4148&(4,13*2,1*3,0,1,0*3,1*3,3),297.7439  \\
            
          60&5&(45,0*3,10),103.5759&(50,0*3,5),162.4867&            
            (52,0*3,3),399.2079  \\
            &10&(13*3,0*5,1,10),139.7965&(15,16,14,0*6,5),170.2333&(16*3,0*6,2),291.614 \\
            &15&(4,5,6,5,6,3,4,1,0,1,0,1*2,2,6),185.1344&(11,5,13,2*3,0,1*2,0,1,0,1*2,5),207.0492&(6,8,10,8,5,1,3,0,1,0*2,1,0*2,2),298.6105  \\
          65&5&(50,0*3,10),103.3568&(55,0*3,5),161.8478& 
            (57,0*3,3),396.833  \\
            &10&(15*3,0*6,10),139.6169&(19*2,11,0*4,1,0,5),169.7746&(18,17,18,0*6,2),292.2875  \\
            &15 &(4,7,6,4,3,5,0,1,2,0*2,1,3,7*2),181.2025&(10,2,12,11,2*2,1,0,1,0,1*4,5),207.4059&(14,3,18,6,0,1*4,0*3,1*2,3),297.4492  \\
         \bottomrule
    \end{tabular}
\end{sidewaystable}

\begin{table}[h] 
    \centering
    \caption{Optimal schemes based on $\phi$ and their relative efficiencies compared to $\phi_{0}$ with $(\kappa_1,\kappa_2,\kappa_3)=(10,50,250)$ and $n=20$,$m=5$ for Weibull distribution with $\rho =1$ and $\zeta=\phi$}
    \label{table:3}
    \vspace{5pt}
    \begin{tabular}{c c c c}
        \toprule
        $\phi_{0}$ & $\phi$    & $(R_{1},R_{2},R_{3},R_{4},R_{5})|\phi$ & $RE_{1}(\phi)$ \\
          \midrule
           2.00  & 1.85 & (6*2,0*2,3) & 1.0000  \\
           2.00  & 1.90 & (6*2,0*2,3) & 1.0000  \\
           2.00  & 1.95 & (6*2,0*2,3) & 1.0000  \\
           2.00  & 2.00 & (6*2,0*2,3) & 1.0000  \\
           2.00  & 2.05 & (6*2,0*2,3) & 1.0000  \\
           2.00  & 2.10 & (6*2,0*2,3) & 1.0000  \\
           2.00  & 2.15 & (6*2,0*2,3) & 1.0000  \\
                 &      &             &         \\
           0.50  & 0.35 & (7*2,0*2,1) & 1.0000  \\
           0.50  & 0.40 & (7*2,0*2,1) & 1.0000  \\
           0.50  & 0.45 & (7*2,0*2,1) & 1.0000  \\
           0.50  & 0.50 & (7*2,0*2,1) & 1.0000  \\
           0.50  & 0.55 & (7*2,0*2,1) & 1.0000  \\
           0.50  & 0.60 & (7*2,0*2,1) & 1.0000  \\
           0.50  & 0.65 & (7*2,0*2,1) & 1.0000  \\
          \bottomrule
      \end{tabular}
\end{table}

\begin{table}[h] 
    \centering
    \caption{Optimal schemes based on $\mathcal{C}$ and their relative efficiencies compared to $\mathcal{C}_{0} =(10,50,250)$ with $n=20$,$m=5$ for Weibull distribution with $\rho =1$ and $\zeta=1$}
    \label{table:4}
    \vspace{5pt}
    \begin{tabular}{c c c c}
        \toprule
        $\mathcal{C}_{0}=(\kappa_1,\kappa_2,\kappa_3)$ & $(R_{1},R_{2},R_{3},R_{4},R_{5})|\mathcal{C}$ & $RE_{2}(\mathcal{C})$ \\
        \midrule
           (7,50,250)   & (7,6,0*2,2) & 1.0000  \\
           (8,50,250)   & (7,6,0*2,2) & 1.0000  \\
           (9,50,250)   & (7,6,0*2,2) & 1.0000  \\
           (11,50,250)  & (7,6,0*2,2) & 1.0000  \\
           (12,50,250)  & (7,6,0*2,2) & 1.0000  \\
           (13,50,250)  & (7,6,0*2,2) & 1.0000  \\
                        &             &         \\
           (10,47,250)  & (7,6,0*2,2) & 1.0000  \\
           (10,48,250)  & (7,6,0*2,2) & 1.0000  \\
           (10,49,250)  & (7,6,0*2,2) & 1.0000  \\
           (10,51,250)  & (7,6,0*2,2) & 1.0000  \\
           (10,52,250)  & (7,6,0*2,2) & 1.0000  \\
           (10,53,250)  & (7,6,0*2,2) & 1.0000  \\
                        &             &         \\
           (10,50,247)  & (7,6,0*2,2) & 1.0000  \\
           (10,50,248)  & (7,6,0*2,2) & 1.0000  \\
           (10,50,249)  & (7,6,0*2,2) & 1.0000  \\
           (10,50,251)  & (7,6,0*2,2) & 1.0000  \\
           (10,50,252)  & (7,6,0*2,2) & 1.0000  \\
           (10,50,253)  & (7,6,0*2,2) & 1.0000  \\
           \bottomrule
    \end{tabular}
\end{table}

\begin{sidewaystable}[h] 
    \centering
    \caption{Optimal progressive censoring schemes for given $n$  with $(\kappa_1,\kappa_2,\kappa_3)=(10,50,250)$}
    \label{table:5}
    \vspace{5pt}
    \begin{tabular}{c c c c}\hline
     & Optimal solution $ \{m^*,\mathcal{R}^*\},\xi(\mathcal{R}^*)$&  &   \\
       \toprule
        $n$  & {Parameters $(\zeta,\rho)$} & {Parameters $(\zeta,\rho)$} & {Parameters $(\zeta,\rho)$}\\
        & (2,1)  (2,0.5)  & (1,1)  (1,0.5) &(0.5,1)  (0.5,0.5) \\
       \midrule
       10  &\{3,(4,0,3)\},103.698& \{5,(1,2,0*2,2)\},178.6414& \{7,(1,0*2,1,0*2,1)\},388.6466  \\
       15 &\{3,(9,0,3)\},102.1896&\{6,(3*2,0*3,3)\},169.3459&\{9,(1*3,2,0*4,1)\},331.9619  \\
       20 &\{3,(14,0,3)\},101.173&\{6,(6*2,0*3,2)\},166.4583&\{11,(1,2*3,1,0*5,1)\},307.8155  \\
       25 &\{3,(19,0,3)\},100.4394&\{6,(8*2,0*3,3)\},164.6452&\{12,(2,3*2,2,1,0*6,2)\},300.091  \\
       30 &\{3,(24,0,3)\},99.89178&\{6,(11*2,0*3,2)\},163.3651&\{12,(4*4,0*7,2)\},296.5222  \\
       35 &\{3,(29,0,3)\},99.46189&\{6,(13*2,0*3,3)\},162.5573&\{12,(5*2,4,5,0*7,4)\},249.4874  \\
       \bottomrule
       \end{tabular}
\end{sidewaystable}
\clearpage
\newpage

%\bibliographystyle{plain}
%\bibliography{reference}
%\bibliography{sn-bibliography-ujjwal}% common bib file

\end{document}